\documentclass[aip,sd,amsmath,amssymb,reprint]{revtex4-1}

\usepackage{graphicx}% Include figure files
\usepackage{dcolumn}% Align table columns on decimal point
\usepackage{bm}% bold math

\begin{document}

%\preprint{AIP/123-QED}

\title{Photoinduced two-step insulator-metal transition in $\rm{Ti_4O_7}$ by ultrafast time-resolved optical reflectivity}

\newcommand{\SUPL}{Strong-field and Ultrafast Photonics Lab, Institute of Laser Engineering,
                                   Beijing University of Technology, Beijing 100124, P. R. China}
\newcommand{\CCE}{Department of Chemistry and Chemical Engineering, College of Environmental and Energy Engineering,
                                   Beijing University of Technology, Beijing 100124, P. R. China}
\newcommand{\CAS}{College of Applied Sciences, Beijing University of Technology, Beijing 100124, P. R. China}
\newcommand{\SUP}{Hunan Key Laboratory of Super-microstructure and Ultrafast Process, School of Physics and Electronics,
                                   Central South University, Changsha, Hunan 410083, P. R. China}
\newcommand{\CSU}{School of Physics and Electronics, Central South University, Changsha, Hunan 410083, P. R. China}

\author{X. C. Nie}
\affiliation{\SUPL}  %\affiliation{\BJUT}
\author{Hai-Ying Song}
\affiliation{\SUPL}
\author{Xiu Zhang}
\affiliation{\SUPL}
\author{Shi-Bing Liu}
\affiliation{\SUPL}
\author{Fan Li}
\affiliation{\CCE}
\author{Lili Yue}
\affiliation{\CAS}
\author{Jian-Qiao Meng}
\affiliation{\SUP}
\author{Yu-Xia Duan}
\affiliation{\CSU}
\author{H. Y. Liu}
   \email{haiyun.liu@bjut.edu.cn}
\affiliation{\SUPL}

\date{\today}% It is always \today, today, but any date may be explicitly specified

\begin{abstract}
We report on systematic investigation of hot carrier dynamics in $\rm{Ti_4O_7}$ by ultrafast time-resolved optical reflectivity.
We find the transient indication for its two-step insulator-metal (I-M) transition, in which two phase transitions occur from long-range order bipolaron low-temperature insulating (LI) phase to disordered bipolaron high-temperature insulating (HI) phase at $T_{c1}$ and to free carrier metallic (M) phase at $T_{c2}$. Our results reveal that photoexcitation can effectively lower down both $T_{c1}$ and $T_{c2}$ with pump fluence increasing, allowing a light-control of I-M transition. We address a phase diagram that provides a framework for the photoinduced I-M transition and helps the potential use of $\rm{Ti_4O_7}$ for photoelectric and thermoelectric devices.

\end{abstract}

% PACS, the Physics and Astronomy Classification Scheme.
       % 71.27.+a {Strongly correlated electron systems; heavy fermions}           %71.38.Mx {Bipolarons}
       %71.30.+h {Metal-insulator transitions and other electronic transitions}      %64.60.Cn {Order-disorder transformations}
       %78.47.+p {Time-resolved optical spectroscopies and other ultrafast optical measurements in condensed matter}
  %Use showkeys class option if keyword display desired

\maketitle
% --------------------------------------------------------------------------------------------------------------------------------
The insulator-metal (I-M) transition has been attracting a lot of interests over the past half century in condensed matter physics\cite{{Mott-IMT},{Mott-IMTs}}, however, the mechanism in many branches still remains controversial\cite{{Shamblin-NC-2018}}. Various scenarios have been proposed, including the opening of a gap due to spectral weight transfer\cite{Chu-NM-2017}, Anderson localization of electrons induced by lattice disorder\cite{{Anderson-PR-1958}, {Ying2016Anderson}}, polaron self-trapping driven by electron-phonon ($e\text{-}ph$) interactions\cite{{Tokura2017Emergent}}, and so forth. In addition to the control of phase transition by equilibrium methods, such as temperature, chemical-doping and pressure variation, photoinduced I-M transition in nonequilibrium states is of great interest for the study of ultrafast physics and the realization of the light-control of electronic properties. For example, in semiconductor GaAs, transient I-M transition can be induced by the interplay between free carriers and excitons, accompanied by a sign reverse of time-resolved reflectivity changes\cite{{Nie-NJP-2018}}. By implementing ultrafast photoexcitation, a growing interest has been devoted to photoinduced I-M transition in strongly correlated systems, through melting of charge-density-wave (CDW) order in low-dimensional CDW materials\cite{Tomeljak-PRL-2009,Petersen-PRL-2011,HY-PRB-2013}, and resonant phonon-driven melting of orbital or magnetic order in magnetoresistive manganites\cite{DP-NM-2007,{Baldini-PRB-2018}}, etc.

Titanium oxide $\rm{Ti_4O_7}$ is a member of transition-metal oxides with Magn$\rm{\acute{e}}$li phase, which hosts a triclinic structure and a characteristic two-step I-M transition accompanied by large reflectivity and heat capacity changes\cite{{Lakkis-PRB-1976},{Watanabe-JLumin-2007},{Taguchi-PRL-2010},{Schlenker-PMB-1980}}. It has one 3$d$ electron per two Ti ions, allowing two possible valence states of Ti$^{3+}$(3$d^1$)
and Ti$^{4+}$(3$d^0$). As depicted in Fig. 1(a), in the high-temperature metallic (M) phase, Ti ions have an uniform
valence state of $+$3.5 and the 3$d$ electrons are delocalized and contribute to the electrical conductivity. Below 150 K ($T_{c2}$), Ti ions take alternative valence state of $\rm{Ti^{3+}}$ and $\rm{Ti^{4+}}$ and the 3$d$ electrons are localized to form covalently bonded $\rm{Ti^{3+}}$-$\rm{Ti^{3+}}$ pairs, which can be considered as bipolarons\cite{{Schlenker-PMB-1980},{Chakraverty-PMB-1980}}. For the low-temperature insulating (LI) phase below 130 K ($T_{c1}$) well ordered $\rm{Ti^{3+}}$-$\rm{Ti^{3+}}$ pairs appear, while this long-range order vanishes in the high-temperature insulating (HI) phase (130 K $<T<$ 150 K)\cite{{Watanabe2006Raman},{Kamioka-JPPA-2015},{Kamioka-APL-2016},{Kobayashi-EPL-2002},{Taguchi-PRL-2010},{Schlenker-PMB-1980},{Chakraverty-PMB-1980},{Zhong2015Electronic},{Weissmann2011Electronic},{Liborio2009Electronic}}. 
Titanium dioxide has been widely used in photocatalysis in rechargeable batteries\cite{{Lee-ACS-2018},{Yao-JMC-2012}}. Much attention has been drawn to resistive switching in titanium dioxide due to it's possible application in nonvolatile memory
devices called Resistance Random Access Memory (ReRAM)\cite{{Ko-APL-2012},{Kwon-NN-2010}}. Recently optical-pump terahertz-probe measurements has observed the photoinduced I-M transition $\rm{Ti_4O_7}$ by monitoring the dynamic relaxation on the timescale of tens of picoseconds (ps)\cite{Kamioka-APL-2016}. However, systematic investigation on the fast dynamics in sub-ps, the photoinduced I-M via two-step procedure and phase diagram, are still lacking.

Ultrafast time-resolved optical reflectivity has been proved to be an effective method to track hot carrier dynamics in
semiconductors\cite{{Nie-NJP-2018}} and correlated electron materials\cite{{Torchinsky2010Band},{Tian2016Ultrafast}}, including cuprate high-$T_c$ superconductors (HTSC)\cite{{Vishik2017Ultrafast},{Torchinsky2013Fluctuating},{Nie-arXiv-2018}} and many other transition-metal oxides\cite{Cilento2010Ultrafast}. Hot carrier dynamics is closely related to the gap characterizations in these materials, where the initially excited hot electron distribution thermalizes through electron-electron ($e\text{-}e$) scattering within tens of femtoseconds (fs), and afterwards a bottleneck effect appears above the gaps accompanied by slower relaxation and cooling to the lattice temperature via the $e\text{-}ph$ scattering.

In this paper we present systematic studies of the hot carrier dynamics in polycrystalline $\rm{Ti_4O_7}$ using ultrafast time-resolved optical reflectively. Our results reveal that, with increasing temperature, the oxide undergoes a two-step phase transition from ordered bipolaron LI
phase to disordered bipolaron HI phase at $T_{c1}$ and to free carrier M phase at $T_{c2}$, accompanied by clear slope changes in the transient amplitude and relaxation curves. In addition, our study clearly shows that both $T_{c1}$ and $T_{c2}$ decrease with the increase of pump fluence, indicating that photoexcitation is an effective way of inducing I-M transition in this system. The possible mechanism of the photoinduced I-M transition has been discussed.

The ultrafast time-resolved optical reflectivity measurements were performed on polycrystalline $\rm{Ti_4O_7}$ using a pump-probe setup, in which optical pulses with temporal duration of 35 fs and centre wavelength of 800 nm, were produced by a regeneratively amplified Ti:sapphire laser system operating at a repetition rate of 1 kHz\cite{{Nie-NJP-2018},{Nie-arXiv-2018}}. The mirror-like sample surface was obtained by mechanical polish. The pump beam was focused to the sample at near-normal incidence with a spot diameter of $\sim$ 0.4 mm and the probe spot diameter was $\sim$ 0.2 mm, measured by imaging the spots in the focal plane of a CCD camera, ensuring an excellent pump-probe overlap. With the pump beam modulated by a chopper, the reflected probe signal was focused onto a Si-based detector which was connected to a lock-in amplifier where the photoinduced transient reflectivity variation $\Delta R/R$ were recorded, with a sensitivity at the level of $10^{-6}$. The temporal evolution of the pump-induced change in the probe reflectivity ($\Delta R/R$) was measured by scanning the delay time
between pump and probe pulses, using a motorized delay line. To perform the temperature-dependent measurements, the sample was mounted
on a cryostat with a temperature sensor embedded close by, allowing a precise control of temperature in the range of 5 $-$ 300 K.
The pump fluence ($F$) was tunable between 70 and 430 $\mu\rm{J/cm^2}$ by using neutral density filters, and the probe fluence of
4 $\mu\rm{J/cm^2}$ was chosen to minimize steady-state heating and maximize the signal-to-noise ratio.

Figure 1 (b) displays time-resolved differential reflectivity $\Delta R/R$ of polycrystalline $\rm{Ti_4O_7}$ as a function of delay time at three
representative temperatures 80 K, 140 K and 180 K, which characterize LI, HI and M phases, respectively. At all temperatures, the photoexcitation
causes a sharp edge around time zero and an abrupt drop at 0.1 ps, corresponding to the very fast thermalization process due to $e$-$e$ scattering, which is hardly to be resolved by our temporal resolution. More interestingly, the following sub-ps recovery shows a strong $T$-dependent behavior: both the amplitude and relaxation increase with temperature decreasing. Obviously, owing to the opening of the bandgap and suppression of any mid-gap states, more hot carriers are accumulated around the bottom of the conduction band right after the thermalization process, resulting in enhanced amplitude and relaxation bottleneck. As indicated by the black curves, the decay is well described by a single-exponential decay function $\Delta R/R=A e^{-t/{\tau}}+A_0$ on a picosecond (ps) timescale, where the amplitude $A$ is proportional to the photoexcited hot carrier density, $\tau$ is the characteristic relaxation, and $A_0$ describes much slower equilibration processes out of the period of the measurement, such as heat diffusion,  corresponding to the long-lived reflectivity plateau as depicted in the inset.

% ---------------------------------------------------------------------------------------
 \begin{figure}[tbp]                                                                                                                                             % Fig.1
   \includegraphics[width=1\columnwidth,angle=0]{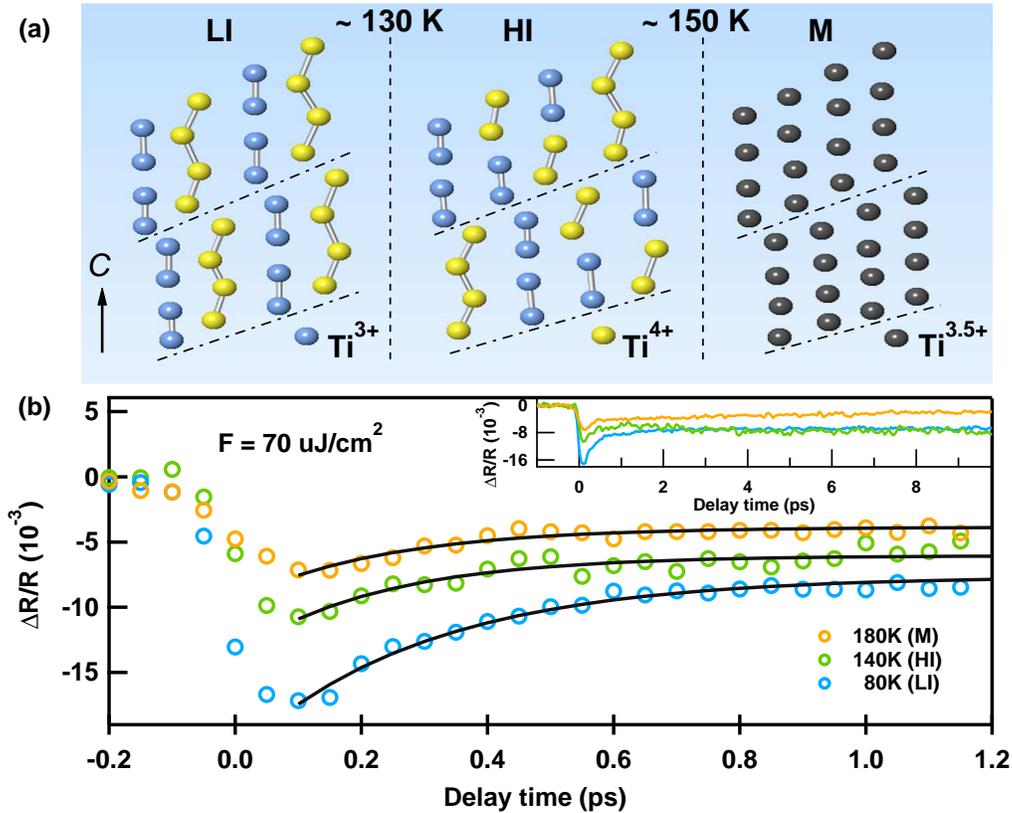}
   \caption{(Color online). (a) Crystal structure of $\rm{Ti_4O_7}$ with the Ti chains parallel to pseudorutile $c$ axis truncated by
   shear planes (chain-dotted lines). The bipolarons ($\rm{Ti^{3+}}$-$\rm{Ti^{3+}}$ pairs) are well ordered in the LI phase, while switch to
   disordered structure in the HI phase. In the M phase, all the Ti ions have an uniform valence of $+$3.5. (b) Time-resolved differential reflectivity 
   $\Delta R/R$ as a function of pump-probe delay time measured at 70 $\mu\rm{J/cm^2}$ at temperatures of 180, 140 and 80 K, respectively.
   The black curves are single-component exponential fits to the recovery. The inset depicts the $\Delta R/R$ curves within 10 ps.
    }
\end{figure}

 % ------------------------------------------------------------------------------------
 \begin{figure}                                                                                                                                                    % Fig.2
   \includegraphics[width=1\columnwidth,angle=0]{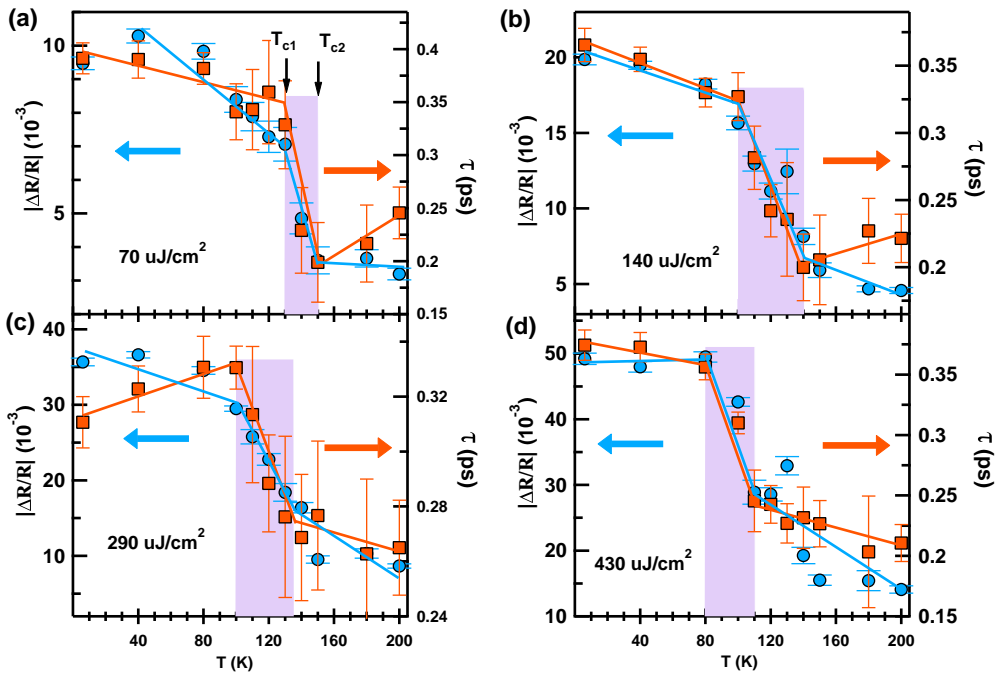}
   \caption{(Color online). Temperature dependence of the amplitudes $A$ (blue circles) and relaxation $\tau$ (orange boxes) extracted by single-exponential fits at pump fluences of 70 (a), 140 (b), 290 (c) and 430 (d) $\mu\rm{J/cm^2}$,
   respectively. The blue and orange lines are used to guide the eye for slop changes. The error bars are obtained from fits. The light red stripes mark the region from the LI-HI phase transition ($T_{c1}$) to the HI-M phase transition ($T_{c2}$).
   }
 \end{figure}
 
 % ------------------------------------------------------------------------------------
\begin{figure}                                                                                                                                                       % Fig.3
\includegraphics[width=1\columnwidth,angle=0]{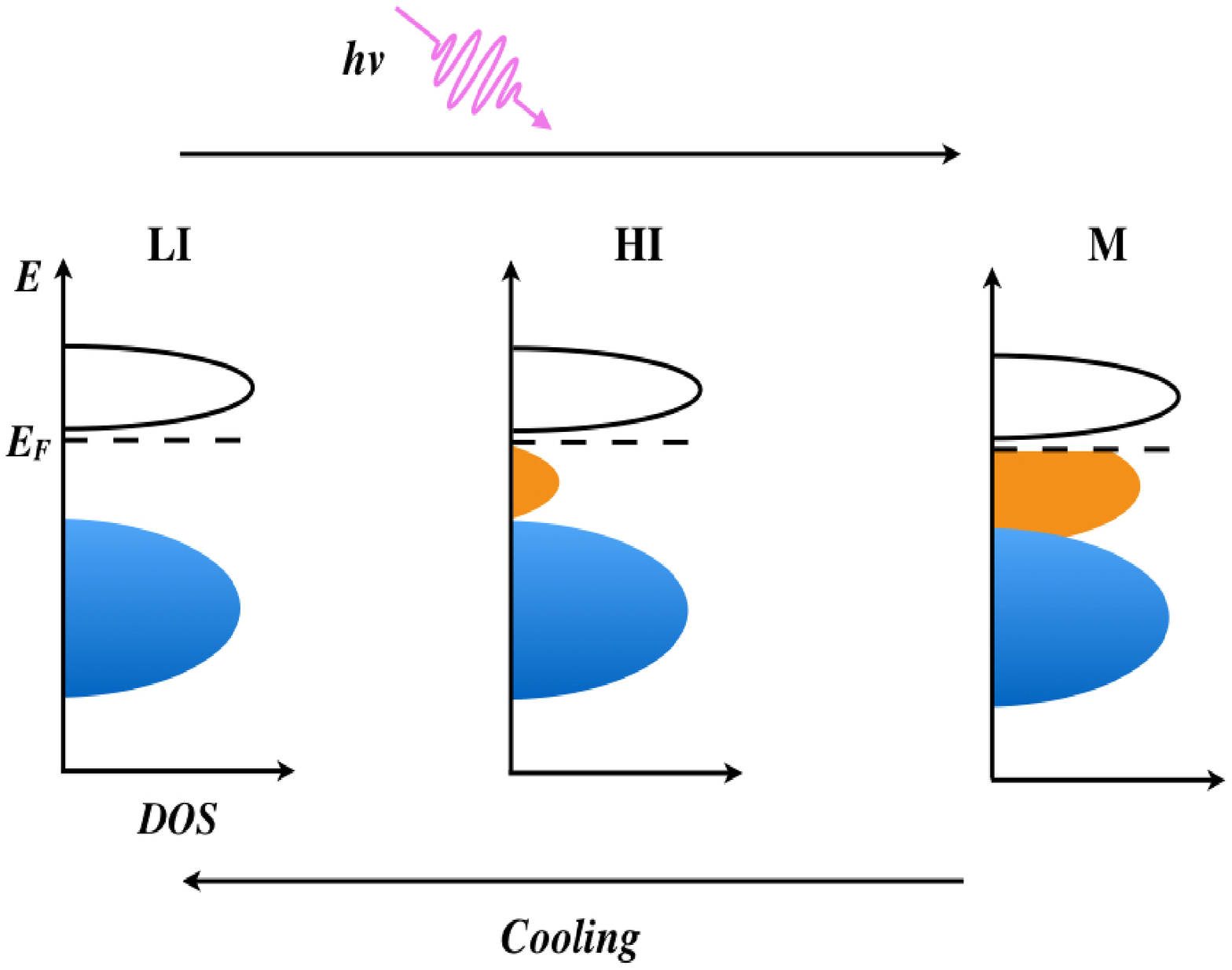}
\caption{(Color online). Schematic evolution of the low energy electronic structure across the two-step I-M transition induced by cooling or photoexcitation. The orange part depicts the Ti 3$d$ contribution to the DOS in the gap. The Fermi level $E_F$ is aligned with the conduction band minimum.
}
\end{figure}
% ------------------------------------------------------------------------------------

In order to quantitatively identify the relation between hot carrier dynamics and phase transitions, in Fig. 2 we plot the amplitudes $A$ and the relaxation times $\tau$ extracted from single-exponential decay fits as a function of temperature at various pump fluences from 70 to 430 $\mu\rm{J/cm^2}$. It is clear that at low pump fluence 70 $\mu\rm{J/cm^2}$ in Fig. 2(a), both the magnitudes $A(T)$ and the relaxation time $\tau(T)$ show slope changes around $T_{c1}$ = 130 K and $T_{c2}$ = 150 K, agrees with the LI-HI and HI-M phase transition temperatures\cite{{Watanabe2006Raman},{Kamioka-JPPA-2015},{Kamioka-APL-2016},{Kobayashi-EPL-2002}}. The light red stripes in the region from $T_{c1}$ to $T_{c2}$ imply that the gap opening might possess a linear relation with temperature. More interestingly, we can see that both $T_{c1}$ and $T_{c2}$ decrease with increasing pump fluence $F$, indicating a significant I-M phase transition by photoexcitation. Upon $F$ = 430 $\mu\rm{J/cm^2}$, $T_{c1}$ and $T_{c2}$ are even as low as 80 and 100 K, as shown in Fig. 2(d).

% --------------------------------------------------------------------------------------------------------------------------------
Figure 3 illustrates the electronic structure associated with the phase transitions. In the free carrier M phase above $T_{c2}$, the delocalization of Ti 3$d$ electrons contributes to the density of states (DOS) around the Fermi level ($E_F$), as well observed both experimentally\cite{{Taguchi-PRL-2010}} and theoretically\cite{{Zhong2015Electronic},{Weissmann2011Electronic},{Liborio2009Electronic}}. In the HI phase ($T_{c1}< T <T_{c2}$), disorganized bipolaron feature forms due to the Anderson localization of the 3$d$ electrons, resulting in suppressed DOS at $E_F$. Then by further cooling to the LI phase $T < T_{c1}$, well ordered bipolarons appear by the inter-site Coulomb interaction, accompanied by the opening of a
insulating gap driven by spectral weight transfer ($\sim$100 meV), as seen by photoemission spectroscopy (PES)\cite{{Kobayashi-EPL-2002},
{Taguchi-PRL-2010}}.

Polaron is a fermionic quasiparticle formed by an electron coupled with a virtual cloud of phonons (distortions), first proposed by Lev Landau in 1933\cite{Landau1933Electron}, for which the electron's mobility lowers and effective mass increases and can be described by the
Fr$\rm{\ddot{o}}$hlich model in the weakly interacting regime\cite{Frohlich1950Properties}. When two polarons are close together, they can further lower their energy by sharing the same distortions, leading to an effective attraction between the polarons. If the interaction is sufficiently large, then that attraction gives rise to a bound bipolaron. As illustrated in Fig. 3, we can interpret our photoexcited I-M transition results as following: photoexcitation generates a large number of free carriers, which can suppress
bipolaron ordering by screening the inter-site Coulomb interaction and force the insulating gap to close, inducing a transition from ordered
bipolaron LI phase to disordered bipolaron HI phase even at lattice temperatures below $T_{c1}$.
By further increasing pump fluence, photoexcitation leads to the dissociation of the bipolarons through the delocalization of Ti 3$d$
electrons, resulting transient metallic properties, analogous to the transition from disordered bipolaron HI phase to free carrier M phase at $T_{c2}$. Therefore, one can precisely control $T_{c1}$ and $T_{c2}$ by careful selection of pump fluence. 
% Delocalization of the Ti 3d electrons leads to metallic properties of the HT phase

\begin{figure}                                                                                                                                                       % Fig.4
   \includegraphics[width=1\columnwidth,angle=0]{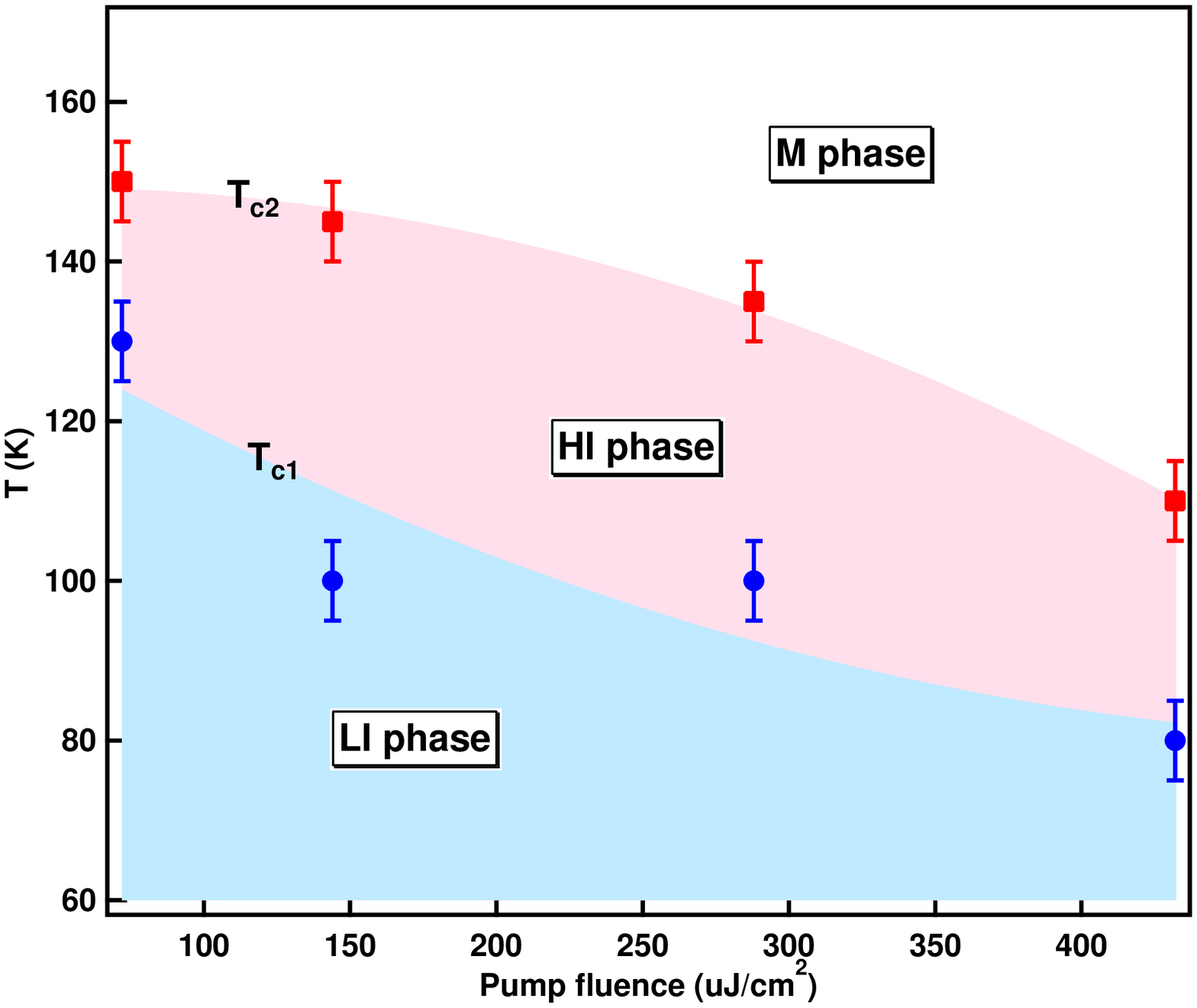}
   \caption{(Color online). Photoexcited phase diagram of $\rm{Ti_4O_7}$, with constant error bars of $\pm$5 K to guide
   the eye. $T_{c1}$ and $T_{c2}$ are the LI-HI and HI-M phase transition temperatures. The color-areas are obtained by polynomial curve fits of $T_{c1}$ and $T_{c2}$.}
\end{figure}

Figure 4 summarizes the phase diagram of $\rm{Ti_4O_7}$ by extracting transition temperatures $T_{c1}$ and $T_{c2}$ from our results
in Fig. 2. The phase diagram consists of three regimes: (1) Above $T_{c2}$, a free carrier metallic state;
(2) Between $T_{c1}$ and $T_{c2}$, a disordered bipolaron HI state and (3) Below $T_{c1}$, a well ordered bipolaron LI state. The $F$ dependence of $T_{c1}$ and $T_{c2}$ allows a light-control of I-M transition.

% -------------------------------------------------------------------------------------------------------------------------------
In summary, we have presented detailed time-resolved optical reflectivity measurements on the polycrystalline titanium oxide $\rm{Ti_4O_7}$.
Our results provide clear evidence for the photoinduced two-step I-M transition from ordered bipolaron LI phase to disordered bipolaron
HI phase and to free carrier M phase. Our investigation indicates that photoexcitation is an effective method to induce
I-M transition and help understanding of the non-equilibrium physics in polaron systems.

% VO2\cite{{Cavalleri2001Femtosecond},Cilento2010Ultrafast}
% -------------------------------------------------------------------------------------------------------------------------------
The authors gratefully acknowledge support from the National Natural Science Foundation of China (Grant No. 51705009) and
NSAF of China (Grant No. U1530153). H. Y. Liu thanks the Sea Poly Project of Beijing Overseas Talents (SPPBOT).

% Bibliography
\bibliographystyle{aipnum4-1}%{abbrv}
\bibliography{Ti4O7}% Produces the bibliography via BibTeX.

\end{document}